# Spatial statistics of superposition of two uncorrelated speckle patterns with polarization diversity

Abhijit Roy

École Polytechnique Fédérale de Lausanne (EPFL), Lausanne 1015, Switzerland

Department of Physics, Indian Institute of Technology Kharagpur, Kharagpur 721302, West Bengal, India

Email: abhijitphy302@gmail.com

A detailed theoretical and experimental study on the effect of the superposition of uncorrelated speckle patterns with polarization diversity on the spatial statistics of the superposed speckle pattern is presented. It is shown that depending on the mutual orientation of the polarization vectors of the constituent speckle patterns, the maximum degree of coherence (DoC) and degree of polarization (DoP) of the superposed speckle pattern changes between a maximum and minimum value in a sinusoidal fashion. Moreover, the average intensity ratio of the constituent speckle patterns is also found to be affecting these variations. A study of the change in the visibility of the two-point intensity correlation function also reveals a sinusoidal nature of the variation and its dependence on the ratio of the average intensity, which are found to be similar to the variations of the maximum DoC and DoP. A detailed study on the changes in the normalized probability density function is also performed for better understanding of the effect on the spatial statistics.

## I. INTRODUCTION

The manipulation and proper characterization of the spatial statistics of speckles, a random intensity distribution, have emerged as a major area of research in recent times due to the wide range of applications of the speckles. The speckles are useful in biomedical studies [1], non-destructive testing and measurement of surface roughness [2], looking through an optical barrier [3] etc. The characterization of the speckles has led to the discovery of the existence of memory effect of a speckle pattern, the implementation of which has made imaging through a scattering medium more convenient [4]. The randomness of spatial intensity distribution of a speckle pattern, which is another important characteristic, can be determined from the study of the Shannon entropy of the speckle pattern [3, 5].

Several other techniques have been developed to measure various parameters, which are required to study different spatial properties of a speckle pattern. The measurement of the spatial degree of coherence (DoC) provides information about the spatial coherency of a speckle pattern and the size of a speckle grain, determined from the intensity correlation of the pattern, is considered as the coherence length of the speckles [3, 6]. Besides, the spatial distribution of polarization can be estimated from the study of the degree of polarization (DoP) [6]. The establishment of a relation between the spatial DoC and DoP has made the characterization of the spatial coherence-polarization (CP) property much easier [7]. An in-depth characterization of the spatial polarization distribution can be performed from the measurement of the generalized Stokes parameters or from the study of the polarization based intensity correlation [8, 9]. The probability density function (PDF) of intensity of a speckle pattern also provides crucial information about the spatial characteristics of a speckle pattern [10].

Although a single speckle pattern can be utilized in understanding the scattering nature of a scattering medium or can be used for imaging applications, the superposition of two or more speckle patterns can offer various advantages over a single speckle pattern. In biomedical studies, the superposition of large number of speckle pattern is utilised to remove the effect of temporal fluctuation of intensity from the measurement [11]. Superposition of many speckle patterns and its temporal averaging can be used for imaging through a scattering medium [12].

Speckle interferometry is another area, where the interference of two speckle patterns from same or different sources is used for various applications. It has been shown that the speckle interferometry can be utilized for non-invasive photoacoustic tomography, where measurement of surface displacement with high resolution is possible using a pulsed laser [13]. In-plane deformation of a microsystem, caused due to various types of perturbations, can also be determined with nanometer accuracy using the speckle interferometry based approach [14].

Another widely used technique based on the speckle interferometry is the off-axis speckle holography, where the speckle interferometry is combined with the two-point intensity correlation of the speckle pattern. The off-axis speckle holography based approach can be used for imaging of 2D objects through a random scattering medium [3]. In the off-axis speckle holography, speckle patterns from two different sources are interfered, and as the speckle patterns are generated from two different sources, they are mutually uncorrelated. Apart from the amplitude imaging, the off-axis speckle holography based approach is also useful for polarization imaging through a random scattering medium [15]. As the speckle superposition has a wide range of applications, it is important to characterize the spatial statistics of the superposed speckle pattern in detail.

It has been reported that the superposition of independent speckle patterns reduces the contrast of the superposed speckle pattern [16]. Manipulation of the contrast can also be achieved, if polarization diversity is introduced to the case of superposition of correlated speckle patterns [17]. Superposition of uncorrelated speckle patterns has been reported to be useful for polarization sensing through a scattering medium [18]. The superposition of two speckle patterns has been reported be affecting the PDF of intensity of the superposed speckle pattern. It has been observed that in case of superposition of fully mutually correlated speckle patterns, the PDF follows an exponential decay behaviour [19]. However, with the decrease of the mutual correlation, the PDF is found be deviating from the exponential decay behaviour [20]. Besides, lot of efforts have also been made to theoretically predict different spatial characteristics, such as, the PDF and average speckle contrast of the superposition of N number of mutually correlated and uncorrelated speckle patterns [21-23]. The two-point intensity correlation and joint-PDF of intensity based approach have also been adopted to characterize the sum of uncorrelated speckle patterns [24]. However, the works reported in Refs. [21-24] are limited to the case of partially developed speckle patterns. In a recent study, the intensity correlation function of fully developed superposed speckle patterns is reported to be affected, if the spatial statistics of any of the constituent fully developed speckle pattern is altered [25]. Most of these reported works are largely confined to the scalar domain. Hence, a detailed study on the effect of the superposition of two uncorrelated speckle patterns with polarization diversity on the spatial statistics of the superposed speckle pattern is needed to be performed.

In this work, polarization diversity is incorporated to the case of superposition of two uncorrelated, fully developed speckle patterns, and a detailed investigation is performed to study the effect on the spatial characteristics of the superposed speckle pattern. The effect on the spatial characteristics is investigated from the study of the change in the maximum DoC, DoP and normalized PDF. As the speckles are used for looking through an optical barrier, where the quality of the retrieved object is determined by the visibility of the two-point intensity correlation function of the speckle pattern, an investigation is also performed to study the effect on the visibility of the two-point intensity correlation function. A detailed theoretical analysis is presented, which is confirmed from the experimental measurements.

**II. THEORETICAL DETAILS**

Let us consider that a scattering layer is illuminated by a linearly polarized field with its polarization vector oriented at a particular angle w.r.t the x-axis. As a scattering layer does not scramble the input polarization information, the random field at the observation plane has the same polarization as that of the input beam. In case of an object beam, if the orientation of the polarization vector is θ w.r.t. the x-axis, the object random field, $\mathbf{E_O}(\mathbf{r}, t)$ at the observation plane can be written as

$$\mathbf{E_O}(\mathbf{r}, t) = E_O(\mathbf{r}, t) \, (\cos\theta \, \hat{\mathbf{x}} + \sin\theta \, \hat{\mathbf{y}}) \tag{1}$$

where $E_O(\mathbf{r}, t)$ is the amplitude of the field, $\mathbf{r}$ is the position vector on the transverse observation plane, t is the time, and $\hat{\mathbf{x}}$ and $\hat{\mathbf{y}}$ are the two mutually orthonormal vectors. Similarly, a reference random field, $\mathbf{E_R}(\mathbf{r}, t)$ with its polarization vector oriented at an angle ϕ w.r.t. the x-axis, can be written as

$$\mathbf{E_R}(\mathbf{r}, t) = E_R(\mathbf{r}, t) \, (\cos\phi \, \hat{\mathbf{x}} + \sin\phi \, \hat{\mathbf{y}}) \tag{2}$$

It should be noted that as the fields under consideration are spatially random and temporarily invariant, the fields are functions of the spatial position vector only. The superposed random field, $\mathbf{E_S}(\mathbf{r})$ is written as

$$\mathbf{E_S}(\mathbf{r}) = (E_O(\mathbf{r})\cos\theta + E_R(\mathbf{r})\cos\phi) \, \hat{\mathbf{x}} + (E_O(\mathbf{r})\sin\theta + E_R(\mathbf{r})\sin\phi) \, \hat{\mathbf{y}} \tag{3}$$

The spatial characteristics of a random field distribution can be investigated from the study of the spatial CP property, the visibility of the two-point intensity correlation function, and the PDF of intensity of the field distribution. The spatial CP property of a random field distribution is studied from the two-point intensity correlation function of the recorded intensity distribution, I(**r**) following Ref. [15] or using the CP matrix, $\Gamma(\mathbf{r_1}, \mathbf{r_2})$ of the random field following Ref. [6]. The spatial DoC, $\gamma(\mathbf{r_1}, \mathbf{r_2})$ can be estimated from the intensity correlation function or the CP matrix as

$$\gamma^2(\mathbf{r_1}, \mathbf{r_2}) = \frac{\langle \Delta I(\mathbf{r_1}) \, \Delta I(\mathbf{r_2}) \rangle}{\langle I(\mathbf{r_1}) \rangle \langle I(\mathbf{r_2}) \rangle} = \frac{\mathrm{tr}[\Gamma(\mathbf{r_1},\mathbf{r_2}) \, \Gamma^\dagger(\mathbf{r_1},\mathbf{r_2})]}{|\mathrm{tr}[\Gamma(0)]|^2} \tag{4}$$

where $\Delta I(\mathbf{r}) = I(\mathbf{r}) - \langle I(\mathbf{r}) \rangle$ is the spatial fluctuation of intensity from its mean value, '⟨.⟩' denotes the ensemble average of the variable, and '*tr*' defines the trace of the matrix. Here, it is important to mention that the average intensity can be estimated from the trace of the CP matrix of the random field at $\mathbf{r_1} = \mathbf{r_2}$ i.e. $\langle I(\mathbf{r}) \rangle = \mathrm{tr}(\Gamma(\mathbf{r},\mathbf{r}))$. The CP matrix of a random field $E(\mathbf{r})$ is written as

$$\Gamma(\mathbf{r_1}, \mathbf{r_2}) = \begin{bmatrix} \langle E_x^*(\mathbf{r_1}) \, E_x(\mathbf{r_2}) \rangle & \langle E_x^*(\mathbf{r_1}) \, E_y(\mathbf{r_2}) \rangle \\ \langle E_y^*(\mathbf{r_1}) \, E_x(\mathbf{r_2}) \rangle & \langle E_y^*(\mathbf{r_1}) \, E_y(\mathbf{r_2}) \rangle \end{bmatrix} \tag{5}$$

The spatial DoP, P(**r**) which provides information about the spatial polarization distribution of a spatially random field, can be determined from the maximum DoC as

$$P^2(\mathbf{r}) = 2\gamma^2(\mathbf{r}, \mathbf{r}) - 1 \tag{6}$$

It can be observed from Eq. (6) that for $\gamma^2(\mathbf{r}, \mathbf{r}) = 1$, the value of P(**r**) is also unity i.e. in this case the spatial polarization distribution is uniform in nature. On the other hand, for $\gamma^2(\mathbf{r}, \mathbf{r}) = 0.5$, the value of P(**r**) is zero, which indicates that in this case, the polarization is distributed randomly over different spatial points. In order to calculate the DoC of the superposed random field using Eq. (4), we exploit the established fact that in case of superposition of two mutually uncorrelated fields, the CP matrix of the superposed random field can be written as the sum of the CP matrices of the constituent random fields under the assumptions that both the random fields follow Gaussian statistics [3, 15]. Hence, the CP matrix of the superposed random field, $\Gamma^S(\mathbf{r_1}, \mathbf{r_2})$ can be written as

$$\Gamma^S(\mathbf{r_1}, \mathbf{r_2}) = \Gamma^O(\mathbf{r_1}, \mathbf{r_2}) + \Gamma^R(\mathbf{r_1}, \mathbf{r_2}) \tag{7}$$

where $\Gamma^O(\mathbf{r_1}, \mathbf{r_2})$ and $\Gamma^R(\mathbf{r_1}, \mathbf{r_2})$ are the CP matrices of the object and reference random field, respectively. As the DoC at $\mathbf{r_1} = \mathbf{r_2}$, which is referred as the maximum DoC, and the DoP, which is estimated from the maximum DoC, provide information about the spatial CP property of a random field distribution, the calculation is focused at $\mathbf{r_1} = \mathbf{r_2}$. The $\Gamma^O$ and $\Gamma^R$ at $\mathbf{r_1} = \mathbf{r_2}$ are derived from Eq. (5) by inserting Eqs. (1) and (2), respectively, and are found to be

$$\Gamma^O(\mathbf{r}, \mathbf{r}) = \langle I_O(\mathbf{r}) \rangle \begin{bmatrix} \cos^2\theta & \sin\theta\cos\theta \\ \sin\theta\cos\theta & \sin^2\theta \end{bmatrix} \tag{8}$$

$$\Gamma^R(\mathbf{r}, \mathbf{r}) = \langle I_R(\mathbf{r}) \rangle \begin{bmatrix} \cos^2\phi & \sin\phi\cos\phi \\ \sin\phi\cos\phi & \sin^2\phi \end{bmatrix} \tag{9}$$

where $\langle I_O(\mathbf{r}) \rangle$ and $\langle I_R(\mathbf{r}) \rangle$ are the average intensities of the object and reference random field, respectively. The $\Gamma^S(\mathbf{r}, \mathbf{r})$ is calculated following Eq. (7) by using Eqs. (8) and (9), and the calculated $\Gamma^S(\mathbf{r}, \mathbf{r})$ is utilized to determine the maximum DoC of the superposed random field following Eq. (4), which is found to be

$$\gamma^2(\mathbf{r}, \mathbf{r}) = \frac{\left(\langle I_O(\mathbf{r}) \rangle \cos^2\theta + \langle I_R(\mathbf{r}) \rangle \cos^2\phi\right)^2 + \left(\langle I_O(\mathbf{r}) \rangle \sin^2\theta + \langle I_R(\mathbf{r}) \rangle \sin^2\phi\right)^2 + 2\left(\langle I_O(\mathbf{r}) \rangle \sin\theta\cos\theta + \langle I_R(\mathbf{r}) \rangle \sin\phi\cos\phi\right)^2}{\left(\langle I_O(\mathbf{r}) \rangle + \langle I_R(\mathbf{r}) \rangle\right)^2} \tag{10}$$

If we define a parameter $m$, which is considered as the ratio of the average intensity of the object random field to that of the reference random field i.e. $m = \frac{\langle I_O(\mathbf{r}) \rangle}{\langle I_R(\mathbf{r}) \rangle}$, the Eq. (10) can be simplified as

$$\gamma^2(\mathbf{r}, \mathbf{r}) = \frac{1 + m^2 + 2m\cos^2(\theta - \phi)}{(1 + m)^2} \tag{11}$$

The DoP is calculated by inserting Eq. (11) into Eq. (6), and is found to be

$$P^2(\mathbf{r}) = \frac{1 + m^2 + 2m\cos 2(\theta - \phi)}{(1 + m)^2} \tag{12}$$

It can be observed from Eqs. (11) and (12) that for a fixed value of $m$, the maximum DoC and the DoP of the superposed random field change with the mutual orientation of the polarization vectors of the constituent random fields in a sinusoidal manner, which indicates the changes in the spatial coherence and spatial polarization distribution of the superposed field distribution. The visibility of the two-point intensity correlation function, $V(\mathbf{r})$ can be estimated from the DoP, $P(\mathbf{r})$ following Ref. [26], as

$$V(\mathbf{r}) = \frac{P(\mathbf{r})^2 + 1}{P(\mathbf{r})^2 + 3} \tag{13}$$

As the visibility of the two-point intensity correlation function depends on the DoP of the field, the visibility is also expected to change with the mutual orientation of the polarization vectors of the constituent random fields in a sinusoidal fashion. The experimental validation of the proposed theory is discussed in detail in the next section.

## III. EXPERIMENTAL DETAILS AND RESULTS

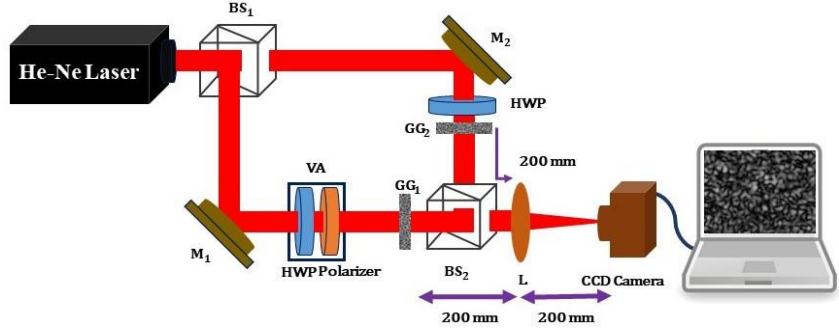

FIG. 1. The schematic diagram of the experimental setup.

The schematic diagram of the experimental setup is shown in Fig. 1. A coherent light beam from a He-Ne laser source with horizontal polarization and a wavelength of 632.8 nm is made to enter a Mach-Zehnder (MZ) interferometer, formed by non-polarizing beam splitters $BS_1$, $BS_2$, and mirrors $M_1$, $M_2$. The beam reflected from $BS_1$ and $M_1$ forms the reference arm of the MZ interferometer, and a ground glass (GG) plate, $GG_1$ is illuminated with this beam. The speckles generated from $GG_1$ are referred as the reference speckles. On the other hand, the beam transmitted through $BS_1$ and folded by $M_2$, which forms the object arm of the MZ interferometer, is passed through another GG plate $GG_2$, and the generated speckles are referred as the object speckles. The object and reference speckles are superposed using $BS_2$ and the far-field superposed speckles are recoded by a CCD camera using a Fourier arrangement constructed using a bi-convex lens, L of focal length of 200 mm. In this Fourier arrangement, the GG plates are kept at the front focal plane of the lens, whereas the CCD camera is placed at the rear focal plane of the lens.

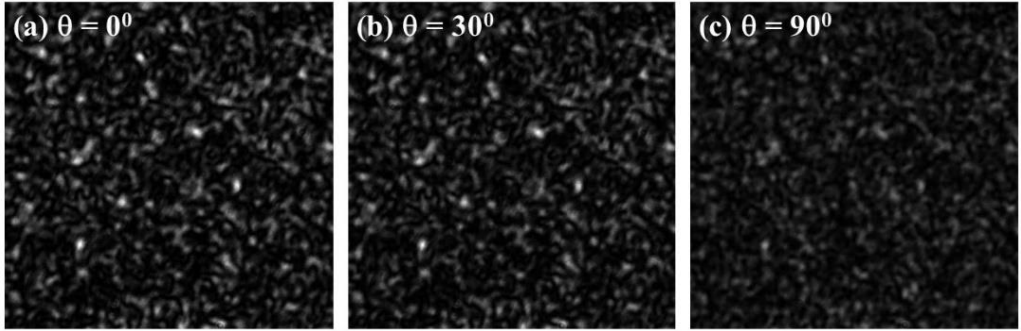

FIG. 2. The recorded superposed speckle patterns for different values of $\theta$.

As the GG plates retain the polarization of the input beam, the reference speckles are horizontally polarized, and hence, in Eq. (11), $\phi = 0^0$. On the other hand, the polarization of the object speckles are controlled using a half-wave plate (HWP), which is placed just before $GG_2$. A variable attenuator, VA, which consists of another HWP and a polarizer is placed in the reference arm of the MZ interferometer to control the average intensity of the reference speckles, i.e. the value of $m$ in Eq. (11). The fast-axis of the HWP is changed from $0^0$ to $180^0$ in steps of $3^0$, and the far-field superposed speckle patterns are recorded in each case. In the present experimental study, apart from $\phi = 0^0$, we have taken m = 1 i.e. the average intensities of the object and reference speckle patterns are same.

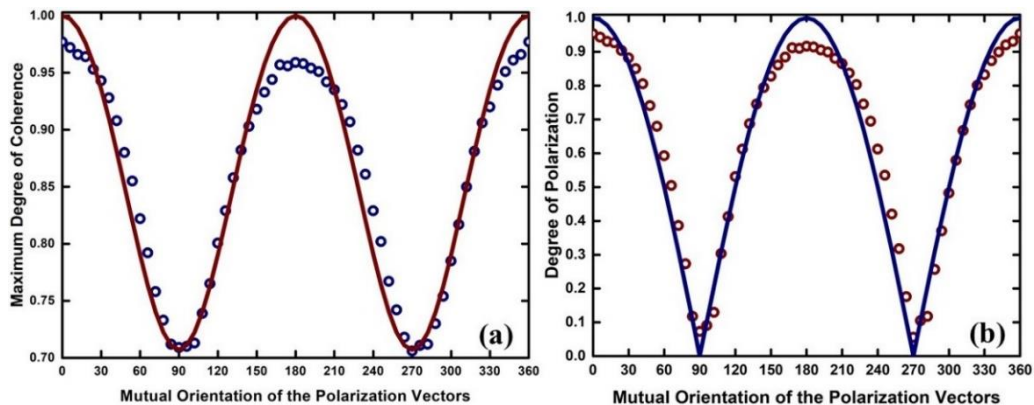

FIG. 3. The theoretical predictions (line) and the experimental observations (circles) of the variations of the (a) maximum DoC and (b) DoP with θ for m = 1.

The recorded speckle patterns for $\theta = 0^0$, $30^0$ and $90^0$ are shown in Fig. 2. The intensity correlation based approach, as discussed in Eq. (4), is exploited to determine the maximum DoC of a recorded superposed speckle pattern by replacing the ensemble average with the spatial average under the assumptions of spatial stationarity and ergodicity of the speckle pattern. The maximum DoC is found to be changing with θ, and the variation is shown in circles in Fig. 3(a), where a sinusoidal variation of the maximum DoC from very close to unity to 0.707 with θ is observed. The theoretical prediction of the variation is also studied following Eq. (11) with $\phi = 0^0$ and m = 1, and is shown as solid line in Fig. 3(a). Similarly, the theoretical prediction (solid line) along with the experimental results (circles) on the variation of the DoP with θ are presented in Fig. 3(b), where the theoretical prediction is based on Eq. (12) and the experimental results are determined following Eq. (6) using the experimentally estimated maximum DoC. It can be observed from Fig. 3(b) that the DoP changes from unity (for $\theta = 0^0$) to zero (for $\theta = 90^0$) with θ in a sinusoidal fashion. The observed variation of the maximum DoC and DoP with θ indicates that the superposition of two mutually uncorrelated speckle patterns with polarization diversity affects the spatial CP property of the superposed speckle pattern, and the spatial polarization distribution of the superposed speckle pattern changes from uniform (P = 1) to random distribution (P = 0) depending on the mutual orientation of the polarization vectors. The experimental results are found to be matching well with the theoretical prediction, although a slight mismatch is observed around the peak, which is due to the imperfection in the used HWP. Although the results presented in Fig. 3 are similar to the results reported in Ref. [27], it is important to mention that the characteristics of the object speckle patterns used in these two studies are completely different. In the present study, the object speckle patterns are fully mutually correlated, whereas in case of Ref. [27], the correlation between the object speckle patterns varies depending on the filtered polarization [9]. Hence, the present study also establishes that modulation of the spatial CP property of the superposed speckle pattern is independent of the mutual correlation of the object speckle patterns.

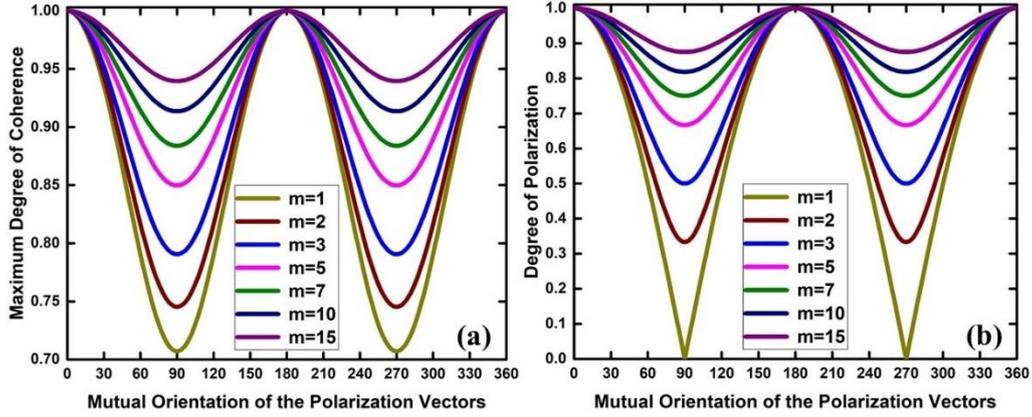

FIG. 4. The theoretically predicted variations of the (a) maximum DoC and (b) DoP with θ for different values of m.

The presence of *m* in Eqs. (11) and (12) indicates that the change in the value of *m* may affect the variations of the maximum DoC and DoP. Therefore, a detailed theoretical investigation on the variation of the maximum DoC and DoP with $\theta$ is performed for different values of *m*, and these variations are presented in Figs. 4(a) and 4(b), respectively, for m = 1, 2, 3, 5, 7, 10, and 15. It can be noticed that with the increase of the value of *m*, although the maximum values of the maximum DoC and DoP remain invariant i.e. unity, the minimum values of these variations start to increase. In case of the variation of the maximum DoC, the minima increases from 0.707 (for m = 1) to very close to 0.95 (for m = 15), whereas in case of the DoP, the minima shifts from zero (for m = 1) to very close to 0.9 (for m = 15). The invariant nature of the maxima is expected, as in these cases, two uncorrelated speckle patterns with the same polarization are superposed, which is not expected to change the polarization of the superposed speckle pattern. Hence, the superposed speckle patterns, in these cases, are uniformly polarized, and the maximum values of the DoP are unity, irrespective of the value of *m*.

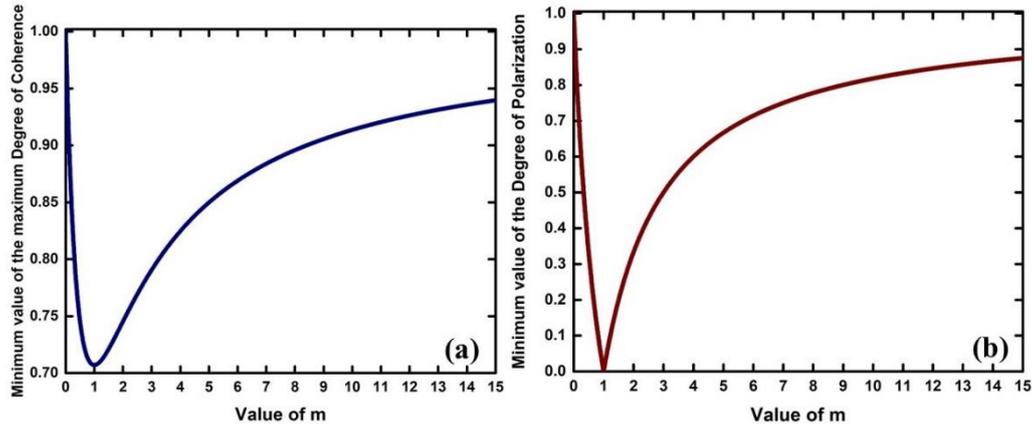

FIG. 5. The change of the minimum value of the (a) maximum DoC and (b) DoP with m.

On the other hand, for m = 1, the theoretically (Figs. 3(b) and 4(b)) and experimentally (Fig. 3(b)) observed minimum values of the DoP are zero, which is similar to the case, where the superposition of two beams with mutually orthogonal polarization and of equal intensity results in a depolarized resultant beam. However, when the value of *m* is more than unity, the minimum value of the DoP is observed to be increasing from zero, as in these cases, one of the two mutually orthogonally polarized speckle patterns has more average intensity than the other one, and the polarization component with higher average intensity dominates over the others, resulting in deviation from the complete random distribution of polarization in the superposed speckle pattern. The change in the minimum of the maximum DoC and DoP with the change of *m* is also studied, and the variations are shown in Figs. 5(a) and 5(b), respectively. It is found that initially, with the increase in the value of *m*, the minimum values of the maximum DoC and DoP decrease sharply from 1.0 and reach to a minimum value of 0.707 and 0, respectively, at m = 1. Subsequently, it start to increase slowly with *m* and reaches very close to 0.95 and 0.9, respectively at m = 15. The variations for m < 1 and m > 1 are expected to be complementary to each other. The maxima of these variations i.e. unity are observed at m = 0 and are expected at m = ∞, as in these two cases, either of the object and reference speckle pattern has zero average intensity. This leads to the presence of either of the uniformly polarized reference and object speckle pattern in the superposed speckles resulting in the maximum possible value of maximum DoC and DoP i.e. unity, which is invariant of the value of θ.

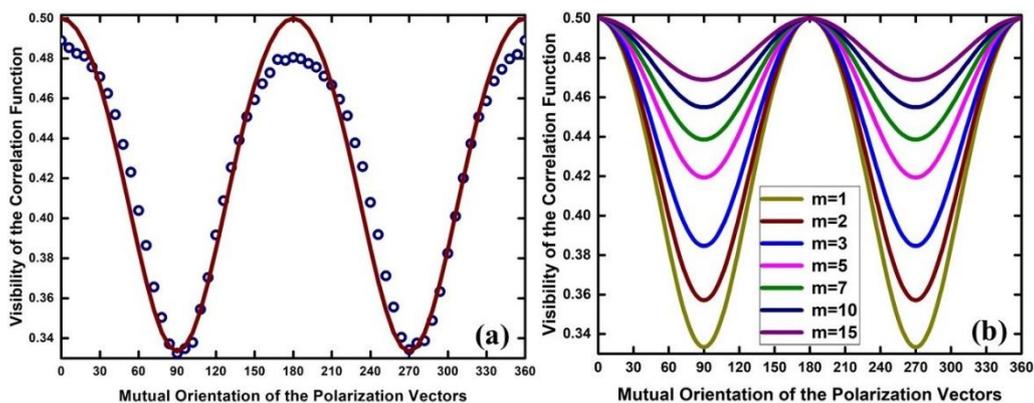

FIG. 6. The variation of the visibility of the intensity correlation function with θ for (a) m = 1 and (b) for different values of m. Circles: the experimental results and line: the theoretical predictions.

The speckles are widely used in different applications, and one of the application is imaging through an optical scattering medium. The two-point intensity correlation based approach is one of the preferred technique for the imaging purpose. The spatial characteristics of a speckle pattern has been reported to play an important role in the object information retrieval process, and it is observed that the visibility of the correlation function of a speckle pattern determines the quality of the retrieved object [15]. The quality of the retrieved object increases with the increase of the visibility of the correlation function and hence, several techniques have been proposed to increase the visibility [28, 29]. Here, in this work, we have characterized the change in the visibility of the two-point intensity correlation function with the mutual orientation of the polarization vectors of the constituent speckle patterns. The experimental variation of the visibility is determined from Eq. (13) using the experimentally measured value of the DoP, whereas the theoretical variation is predicted by inserting Eq. (12) into Eq. (13). The experimental results along with the theoretical prediction for m = 1 are shown in Fig. 6(a), where a close match between these two results are observed. The change in the variation of the visibility for different values of *m* are also calculated and are shown in Fig. 6(b) for the earlier mentioned values of *m*. Similar to the variations of the maximum DoC and DoP, here also, it is observed that the maximum value of the visibility remains unchanged at 0.5, whereas the minimum value of the variation increases from 0.33 to 0.47 with the increase of the value of *m*.

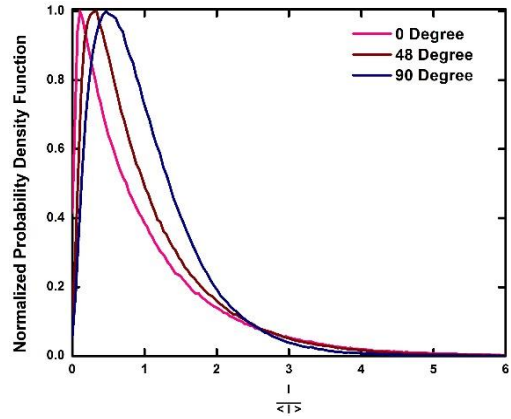

FIG. 7. The normalized PDF of intensity for different values of θ.

It has been observed from the results presented in Figs. 3 and 4 that with the change of the mutual orientation of the polarization vectors of two constituent uncorrelated speckle patterns, the spatial CP property of the superposed speckle pattern changes. As it has been reported that change in the spatial CP property affects the PDF of intensity of a speckle pattern [10], it is expected that a similar observation can also be made in the present study. The normalized PDF of intensity of the superposed speckle patterns for $\theta = 0^0$, $48^0$ and $90^0$ are presented in Fig. 7, where a change in the normalized PDF can be clearly observed. As the shape of the normalized PDF changes with the mutual orientation of the polarization vectors, the change in the area under the normalized PDF is also studied, and the variation is shown in Fig. 8(a). A sinusoidal variation of the area under the normalized PDF with the mutual orientation of the polarization vectors can be observed. It is also found that the area is more at the orientations, where the observed values of the maximum DoC and DoP are at minimum and vice-versa. The change of the area with the experimentally measured DoP is also studied and is presented in Fig. 8(b), where it can be seen that the area reduces slowly with the increase of the DoP. The maximum and minimum of the area are observed to be found at the lowest and highest values of the DoP, respectively.

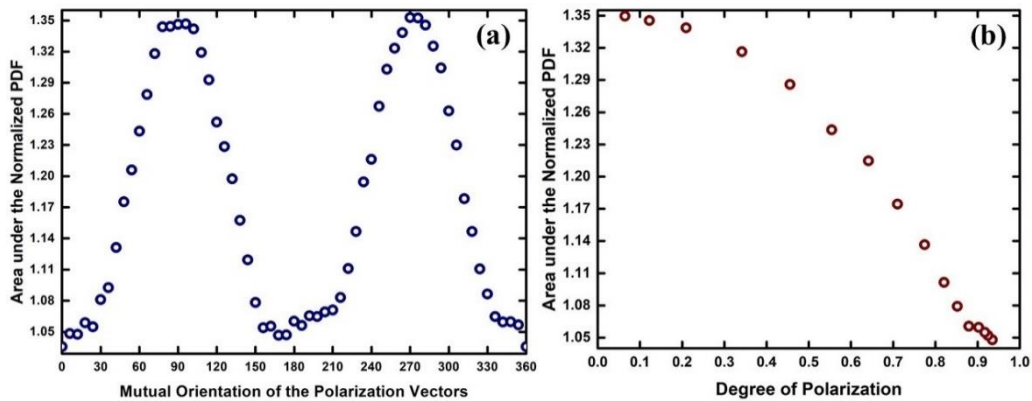

FIG. 8. The variation of the area under the normalized PDF of intensity with: (a) θ and (b) DoP.

A detailed study of Fig. 7 reveals that the magnitude of the normalized PDF at zero intensity also changes with the mutual orientation of the polarization vectors. The change in the magnitude of the normalized PDF at zero intensity with the mutual orientation of the polarization vectors is studied and the variation is presented in Fig. 9(a), where a sinusoidal nature of the variation is observed. The maxima and minima of this variation is found exactly at the same mutual orientations as that of the maxima and minima of the variations of the maximum DoC and DoP. However, in this case, a sharp variation is observed compared to that of the maximum DoC and DoP. The magnitude of the normalized PDF at zero intensity decreases sharply for change in the mutual orientation from $0^0$ to around $50^0$, and subsequently, a slower variation is observed. A similar observation can also be made from the study of the variation of the magnitude as a function of the DoP, which is shown in Fig. 9(b). A slow variation is found for the values of DoP between 0 and 0.7, and subsequently, a sharp increase in the magnitude with the DoP is observed.

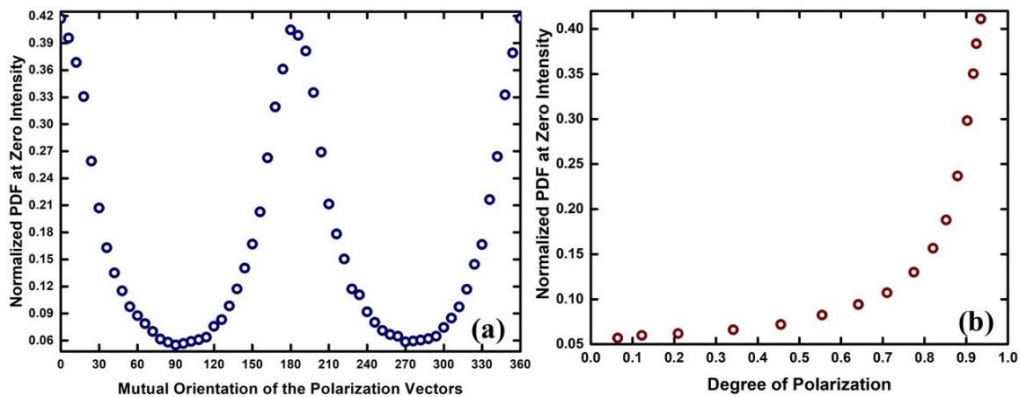

FIG. 9. The variation of the value of normalized PDF at zero intensity with: (a) θ and (b) DoP.

## IV. CONCLUSION

In this paper, we have studied the effect of the superposition of two uncorrelated speckle patterns with polarization diversity and with different spatial average intensity on the spatial statistics of the superposed speckle pattern. A detailed theoretical study has been presented and the theoretically predicted effects are validated experimentally in certain cases. The spatial polarization distribution of the superposed speckle pattern is found to be varying from a uniformly distributed case to a randomly distributed one with the modulation of the polarization diversity of the constituent speckle patterns. The average intensity ratio of the constituent speckle patterns is observed to play an important role in defining the spatial CP property of the superposed speckle pattern, and a change in the ratio along with the polarization diversity are found to have a tremendous effect on the spatial CP property. The change in the visibility of the two-point intensity correlation function is also investigated, which may be useful for imaging applications. The effect of the polarization diversity on the normalized PDF is studied in great detail. The presented detailed study on the change in the area of the normalized PDF and the magnitude of the normalized PDF at zero intensity with the polarization diversity as well as the DoP may provide more insight into the change in the spatial statistics of the superposed speckle pattern. This study may be helpful in different sensing applications and in biomedical studies, where the superposition of speckle patterns with polarization diversity plays an important role.

## ACKNOWLEDGMENT

The author thanks Dr. Rakesh K. Singh from Indian Institute of Technology (BHU) Varanasi, India for his suggestions. The data used in the manuscript were taken in the laboratory of Dr. M. M. Brundavanam at Indian Institute of Technology Kharagpur, India.